\documentstyle[11pt,mrs2001,epsfig]{article}
\begin{document}
\title{AGNS AND CLUSTERS IN CHANDRA DEEP FIELDS}

\author{P. TOZZI$^1$ AND THE CDFS TEAM}
\affil{$^1$Osservatorio Astronomico di Trieste, via G.B. Tiepolo 11,
I--34131, Trieste, Italy}  

\begin{abstract}
The Chandra X--ray Satellite already observed several deep fields,
including the two 1 Megasec exposures of the Chandra Deep Field South
(CDFS) and North.  We review here the main findings from the CDFS.
The LogN--LogS relations show the resolution of the X--ray background
into point sources at the level of 83--99\% in the 1--2 keV band and
65--98\% in the 2--10 keV band, given the uncertainties in the
unresolved value.  The so called ``spectral paradox'' is solved by a
hard, faint population of sources constituted mostly by nearby ($z\leq
1$) absorbed (Type II) AGNs with hard--band luminosities $L\simeq
10^{42}-10^{44}$ erg s$^{-1}$.  When comparing these results to other
deep fields in the X--ray band, we find that the AGNs detected in 0.1
deg$^2$ of the CDFS are representative of the AGNs population as a
whole.  However, we also noticed an excess in the hard counts in two
Chandra deep fields.  If we include this excess and average it among
the observed fields, the total contribution of the XRB can grow of
about 7\%.  Finally, we discuss briefly the properties of the Intra
Cluster Medium imaged at high z, showing no evolution in clusters
properties up to $z\simeq 1$.
\end{abstract}

\section{Introduction}

In the 940 ks exposure (hereafter 1Ms) of the Chandra Deep Field
South (Giacconi et al. 2001; Tozzi et al. 2001; Rosati et al. 2001,
hereafter Paper I, II and III) we reached an on--axis flux limit of
$S=4.5 \times 10^{-16}$ erg s$^{-1}$ cm$^{-2}$ and $S=5.5 \times
10^{-17}$ erg s$^{-1}$ cm$^{-2}$ in the hard (2--10 keV) and soft
(0.5--2 keV) band respectively.  This is the deepest X--ray exposure
to date together with the 1 Megasec exposure reached in the Hubble
Deep Field North by the Penn State University group (see Hornschemeier
et al. 2000; 2001; Brandt et al. 2000).  These, and other deep fields
of few $\times 10^5$ sec, are now public in the Chandra Archive.  To
these deep looks into the X--ray sky, we can add the 100 ksec
observation of the Lockman Hole with XMM (Hasinger et al. 2001).

In these Proceedings we summarize the results from the X--ray data of
the CDFS, namely the soft and hard counts, the contribution to the
X--ray background (XRB), and some X-ray spectral properties of the
sample.  We will also compare the results of the CDFS with other deep
fields.  Finally, we will briefly discuss the properties of the Intra
Cluster Medium (ICM) detected in high--$z$ clusters of galaxies.  This
is just a quick look through the window opened by Chandra and XMM on
the high--z, X--ray sky.

\section{Results from the Chandra Deep Field South}

The Chandra Deep Field South (CDFS) data have been obtained by adding
11 pointings of the Chandra satellite from October 1999 to December
2000.  The pointings had different roll angles, enabling a total
coverage of 0.1089 deg$^2$. The data were reduced with the standard
software developed at the CfA (CIAO release V1.5, see
http://asc.harvard.edu/cda) and the source detection process is
described in detail in Papers I and II.  The energy fluxes are
obtained from the observed net count--rates using a count--rate to
flux conversion factors for an average photon index $\Gamma = 1.4$ and
a Galactic absorbing column of $8 \times 10^{19}$ cm$^{-2}$.  The
conversion factors are $(4.6\pm 0.1) \times 10^{-12}$ erg s$^{-1}$
cm$^{-2}$ per count s$^{-1}$, and $(2.9\pm 0.3) \times 10^{-11}$ erg
s$^{-1}$ cm$^{-2}$ per count $\rm s^{-1}$ in the soft and in the hard
band respectively.  The source count rates are corrected for
vignetting and varying exposure time across the field.

\begin{figure}
\plottwo{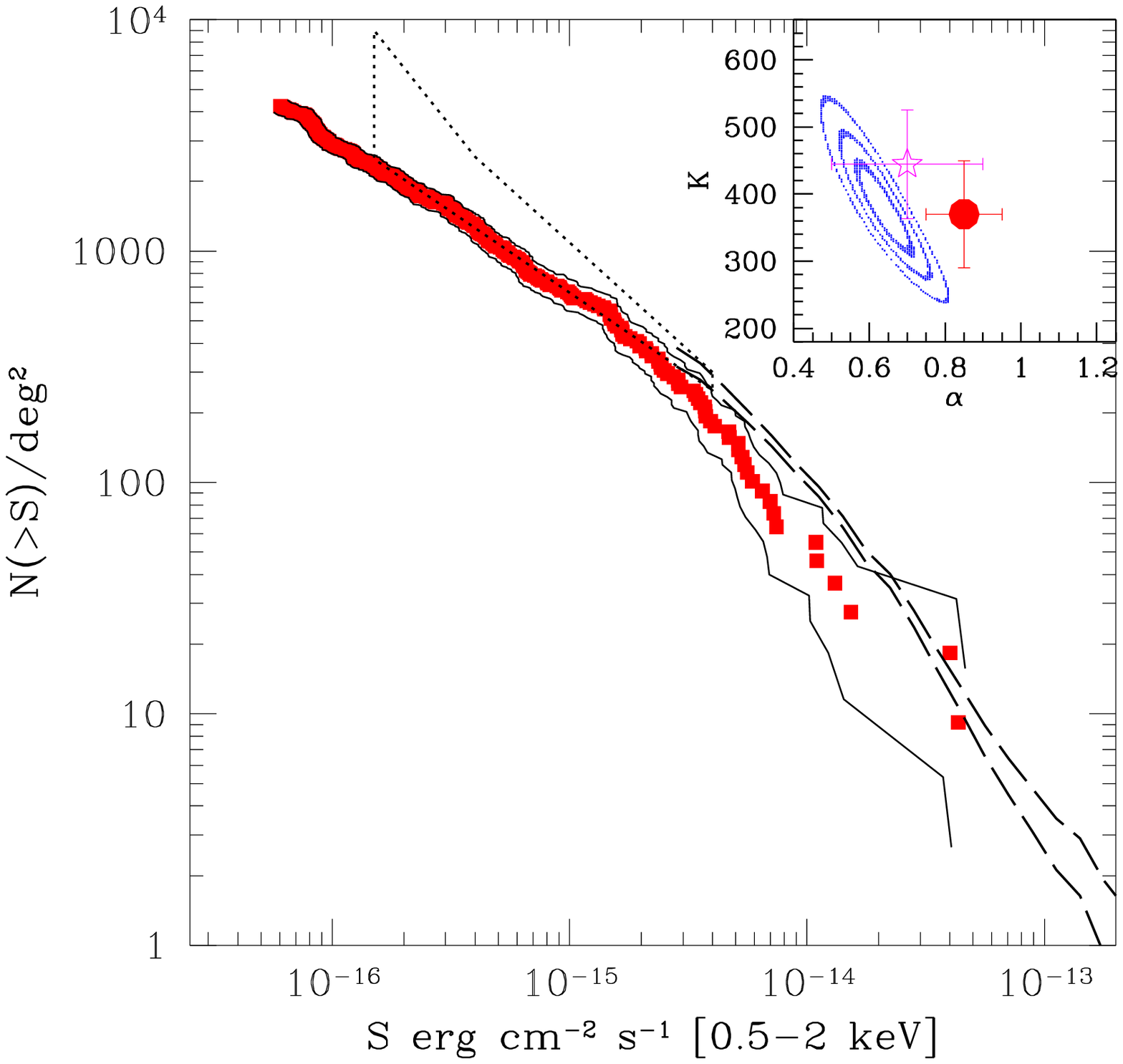}{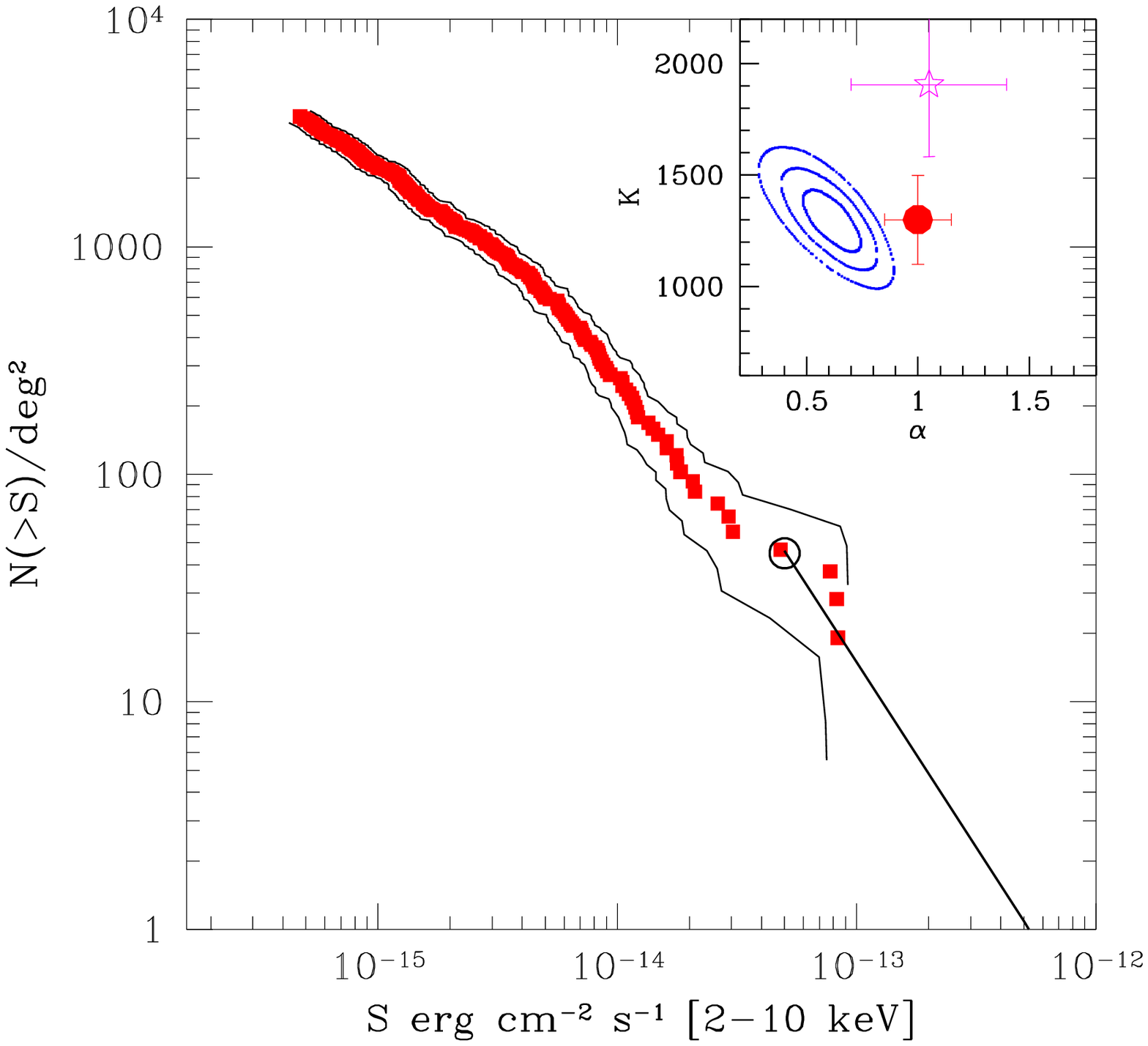}
\caption {Left: The LogN--LogS in the soft band from the CDFS for an
average spectral slope $\Gamma = 1.4$ (filled squares).  Dashed lines
are the counts from the Lockman Hole (Hasinger et al. 1998), and the
dotted contour is the extrapolation from the fluctuation analysis in
ROSAT data (Hasinger et al. 1993).  The upper and lower solid lines
indicate uncertainties due to the sum of the Poisson noise (1 sigma)
including the uncertainty in the conversion factor.  The insert shows
the maximum likelihood contours to the slope and normalization of the
faint end of the number counts from the double power law fit (the
normalization is defined at $S=2\times 10^{-15}$ erg s$^{-1}$
cm$^{-2}$).  The contours correspond to $1 \sigma$, $2 \sigma$ and $3
\sigma$.  The star is the single power law fit from Mushotzky et
al. (2000) at $S=2 \times 10^{-15}$ erg s$^{-1}$ cm$^{-2}$; the error
bar is their quoted 68\% confidence level.  The large dot is the
single power law fit from Paper I.  Right: The LogN--LogS in the hard
band.  The open circle at high fluxes is from ASCA and Beppo SAX
(Giommi, Perri \& Fiore 2000; Ueda et al. 1999) and the solid line is
the fit to the counts from ASCA in the range $10^{-12}-10^{-13}$ erg
cm$^{-2}$ s$^{-1}$ (Della Ceca et al. 2000).
\label{fig1}}
\end{figure}

In the soft band the differential counts are well fitted by a double
power law which is consistent with the Euclidean slope at the bright
end and with a slope $\alpha_{diff} \equiv \alpha+1 = 1.63 \pm 0.13$
(1 sigma error) at the faint end, with a break at $S \simeq 1.5\times
10^{-14}$ erg s$^{-1}$ cm$^{-2}$ (see Paper III).  Thus, below
$S\sim 10^{-15}$ erg s$^{-1}$ cm$^{-2}$, the slope of the cumulative
number counts is $\alpha \simeq 0.6$ (see Figure \ref{fig1}, left),
showing that we are about to saturate the XRB.  In the 1--2 keV band
we find a resolved contribution of $\simeq 6.25 \times 10^{-13} $ erg
s$^{-1}$ cm$^{-2}$ deg$^{-2}$ for fluxes lower than $10^{-15}$ erg
s$^{-1}$ cm$^{-2}$, corresponding to $\simeq 14-17$\% of the
unresolved flux measured by ROSAT (which is $4.4\times 10^{-12}$ erg
s$^{-1}$ cm$^{-2}$, see Hasinger et al. 1998).  If this value is
summed to the contribution at higher fluxes, we estimate a total
contribution of $\simeq 3.65 \times 10^{-12} $ erg s$^{-1}$ cm$^{-2}$
deg$^{-2}$ for fluxes larger than $3 \times 10^{-17}$ erg s$^{-1}$
cm$^{-2}$ (which is our flux limit in the 1--2 keV band),
corresponding to 83\% of the unresolved value.

\begin{figure}
\plottwo{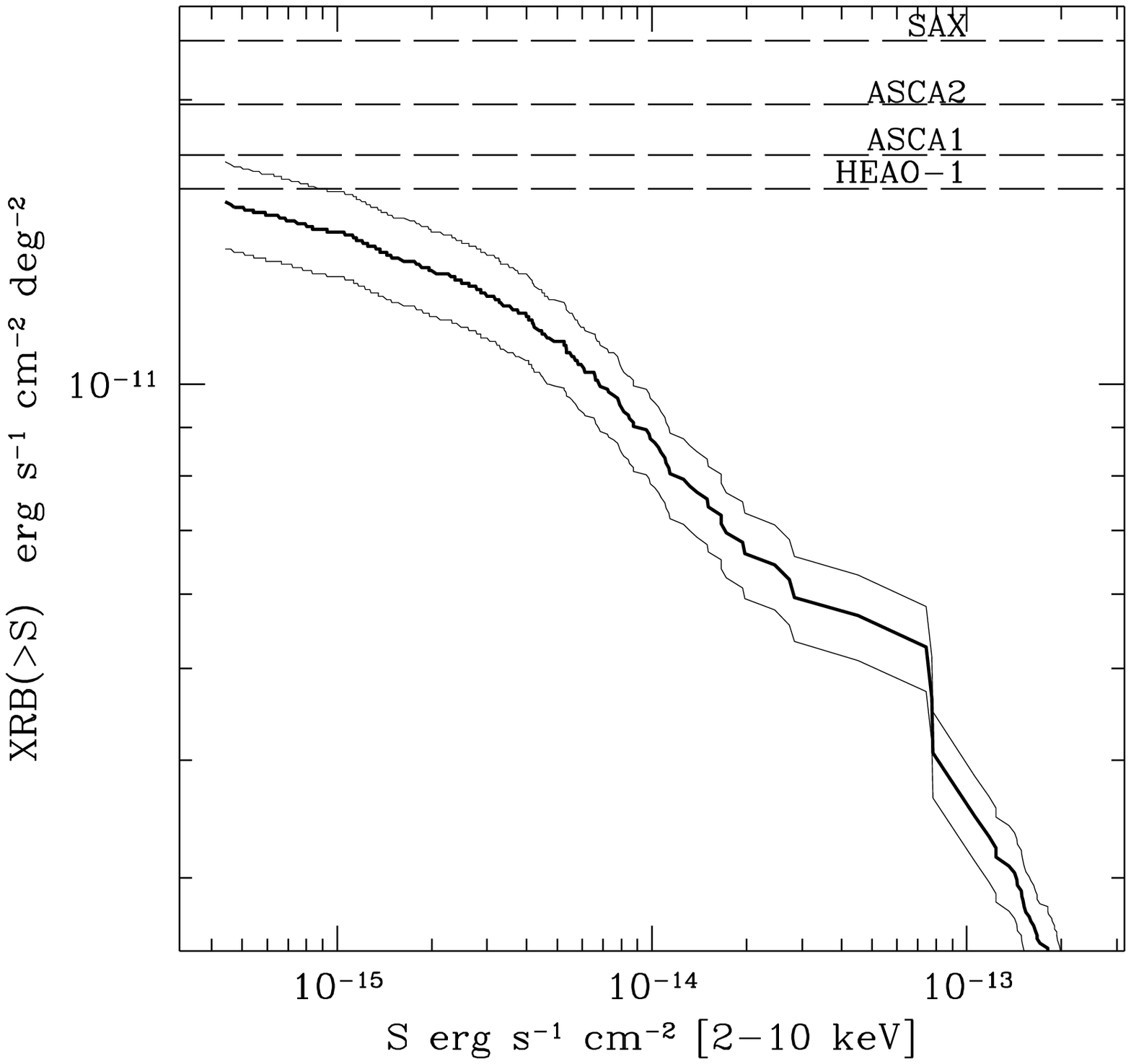}{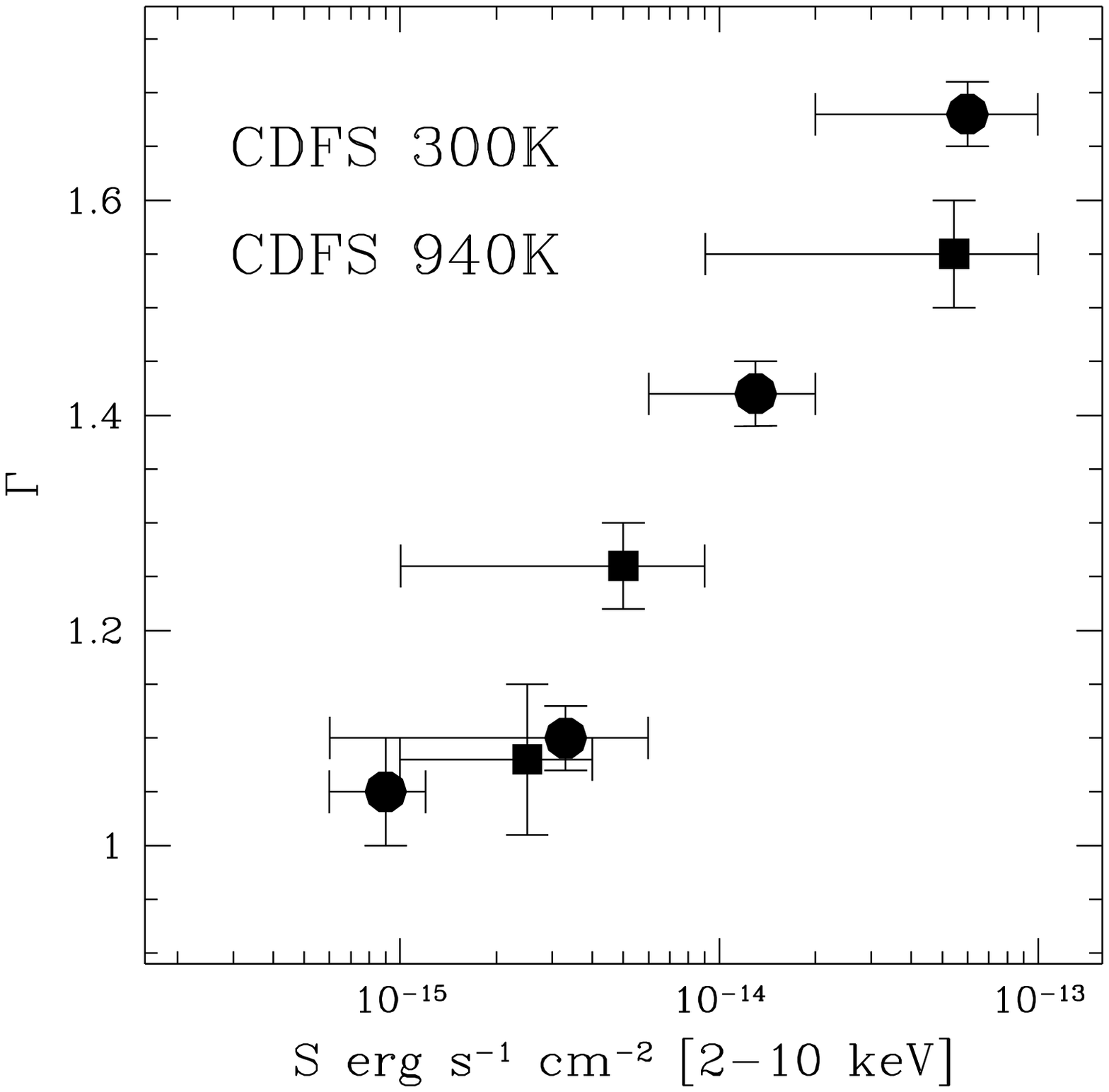}
\caption{Left: The contribution to the hard X--ray flux density as a
function of the flux of the resolved sources (see Rosati et al. 2001).
The total resolved contribution is computed from the 1Ms CDFS sample
plus the bright sample from ASCA at fluxes larger than $\simeq
10^{-13}$ erg s$^{-1}$ cm$^{-2}$ (Della Ceca et al. 2000).  The upper
dashed lines refer to previous measures of the unresolved hard X--ray
background; from bottom to top: Marshall et al. (1980, HEAO-1), Ueda
et al. (1999, ASCA1), Ishisaki (1999, ASCA2), Vecchi et al. (1999,
BeppoSAX).  Right: The average power law index of the stacked spectra
of the bright, medium, faint and very faint subsamples of the sources
detected in the hard band.  Errors on $\Gamma$ refer to the 90\%
confidence level.  The local absorption has been fixed to the Galactic
value $N_H = 8\times 10^{19}$ cm$^{-2}$.  The squares refer to the 300
ks exposure (Paper II), while the circles to the 1Ms exposure.
\label{fig2}}
\end{figure}

In the hard counts too, we find evidence for progressive flattening:
the differential counts are well fitted by a power law which is
consistent with the Euclidean slope at the bright end and with a slope
$\alpha_{diff} = 1.58 \pm 0.12$ (1 sigma) at the faint end, with a
break at $S \simeq 8.4\times 10^{-15}$ erg s$^{-1}$ cm$^{-2}$ (see
Figure \ref{fig1}, right).  The integrated contribution of all the
sources within the flux range $10^{-13}$ erg s$^{-1}$ cm$^{-2}$ to $
4.5 \times 10^{-16}$ erg s$^{-1}$ cm$^{-2}$ in the 2--10 keV band is
$(1.31 \pm 0.20) \times 10^{-11}$ erg s$^{-1}$ cm$^{-2}$ deg$^{-2}$
for $\Gamma=1.4$.  After including the bright end seen by ASCA for $S>
10^{-13}$ erg s$^{-1}$ cm$^{-2}$ (Della Ceca et al. 2000), the total
resolved hard X--ray background amounts to $(1.56 \pm 0.26) \times
10^{-11}$ erg s$^{-1}$ cm$^{-2}$ deg$^{-2}$.  In Figure \ref{fig2}
(left) we show the total contribution computed from the CDFS plus ASCA
sample.  The value at the limiting flux has now reached the value of
the unresolved XRB $1.6 \times 10^{-11}$ erg s$^{-1}$ cm$^{-2}$
deg$^{-2}$ from UHURU and HEAO-1 (Marshall et al. 1980).  More recent
values of the 2--10 keV integrated flux from the BeppoSAX and ASCA
surveys appear higher by 20-40\%.  Thus a fraction between 0.98 and
0.65 of the total XRB flux value has now been resolved into discrete
sources.

The stacked spectrum of the sources of the total sample is well fitted
by a power law with a photon index $\Gamma = 1.375 \pm 0.015$ (errors
refer to the 90\% confidence level) with $\chi^2_{\nu}=1.65$ for a
column density fixed to the Galactic value $N_H = 8 \times 10^{19}$
cm$^{-2}$.  The average spectrum is consistent with the shape of the
unresolved hard background $\langle \Gamma \rangle \simeq 1.4$,
confirming previous findings.  However, we notice that the average
spectrum of the sources is harder at low fluxes.  To investigate this
behaviour, we divided the sample of sources detected in the hard band
in 4 subsamples: bright ($S> 2\times 10^{-14}$ erg s$^{-1}$
cm$^{-2}$), medium ($ 2\times 10^{-14}>S> 6\times 10^{-15}$ erg
s$^{-1}$ cm$^{-2}$), faint ($ S< 6\times 10^{-15}$ erg s$^{-1}$
cm$^{-2}$), and very faint ($ S< 2\times 10^{-15}$ erg s$^{-1}$
cm$^{-2}$).  The average slope of the stacked spectra is $\Gamma =
1.68\pm 0.03$, $1.42\pm 0.03$, $1.10\pm 0.03$, and $1.05\pm 0.05$
respectively (Figure \ref{fig2}, right).  These data show the
emergence of the hard population at faint fluxes, which resolves the
so called {\sl spectral paradox}.  In fact, the spectrum of the
sources detected from previous X--ray missions (namely ROSAT and ASCA)
at fluxes brighter than $10^{-13}$ erg s$^{-1}$ cm$^{-2}$, showed a
spectral slope of $\Gamma \simeq 1.7-1.8$.  Chandra and XMM have now
made possible the direct detection of a population, of absorbed
(TypeII) AGNs with hard--band luminosities in the range $L = 10^{42} -
10^{44}$ erg s$^{-1}$, building up the flat spectrum ($\Gamma = 1.4$)
of the XRB.  Softer sources appear to span the range from $10^{40}$ to
$10^{45}$ erg s$^{-1}$.  At luminosities larger than $10^{43}$ erg
s$^{-1}$ in the soft band, these sources are mostly identified with
TypeI AGN.  At the low luminosity end, several sources are detected
only in the soft band and are identified with normal galaxies at
redshifts less than $0.6$, with luminosities restricted to $L_x
=10^{40}-10^{42}$ erg s$^{-1}$ (see also Hornschemeier et al. 2001).
Such luminosities may be due to a combination of low--level nuclear
activity, a population of low mass X--ray binaries, and thermal
emission from hot halos.

\section{Comparison with other deep fields}

\begin{figure}
\plottwo{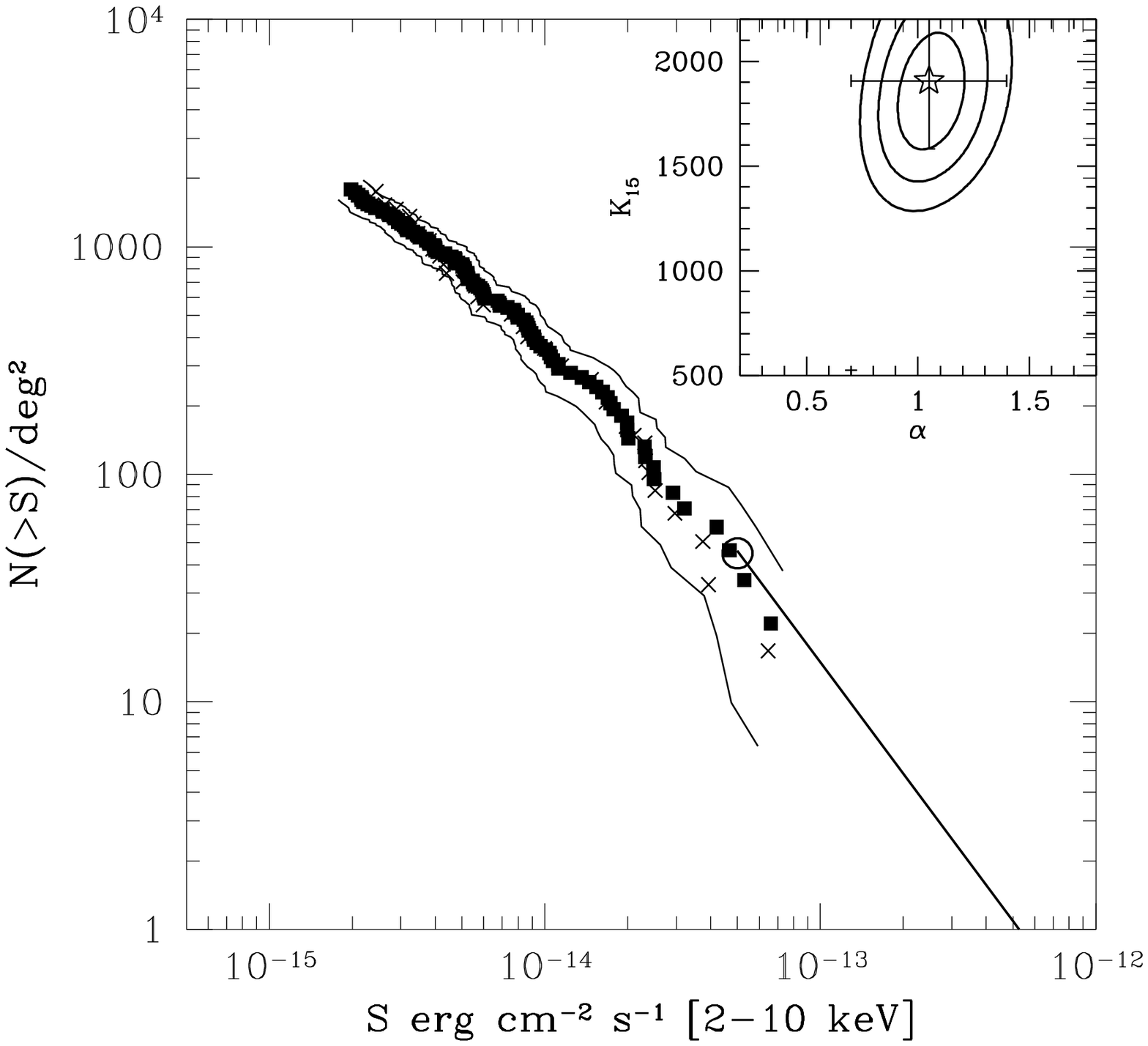}{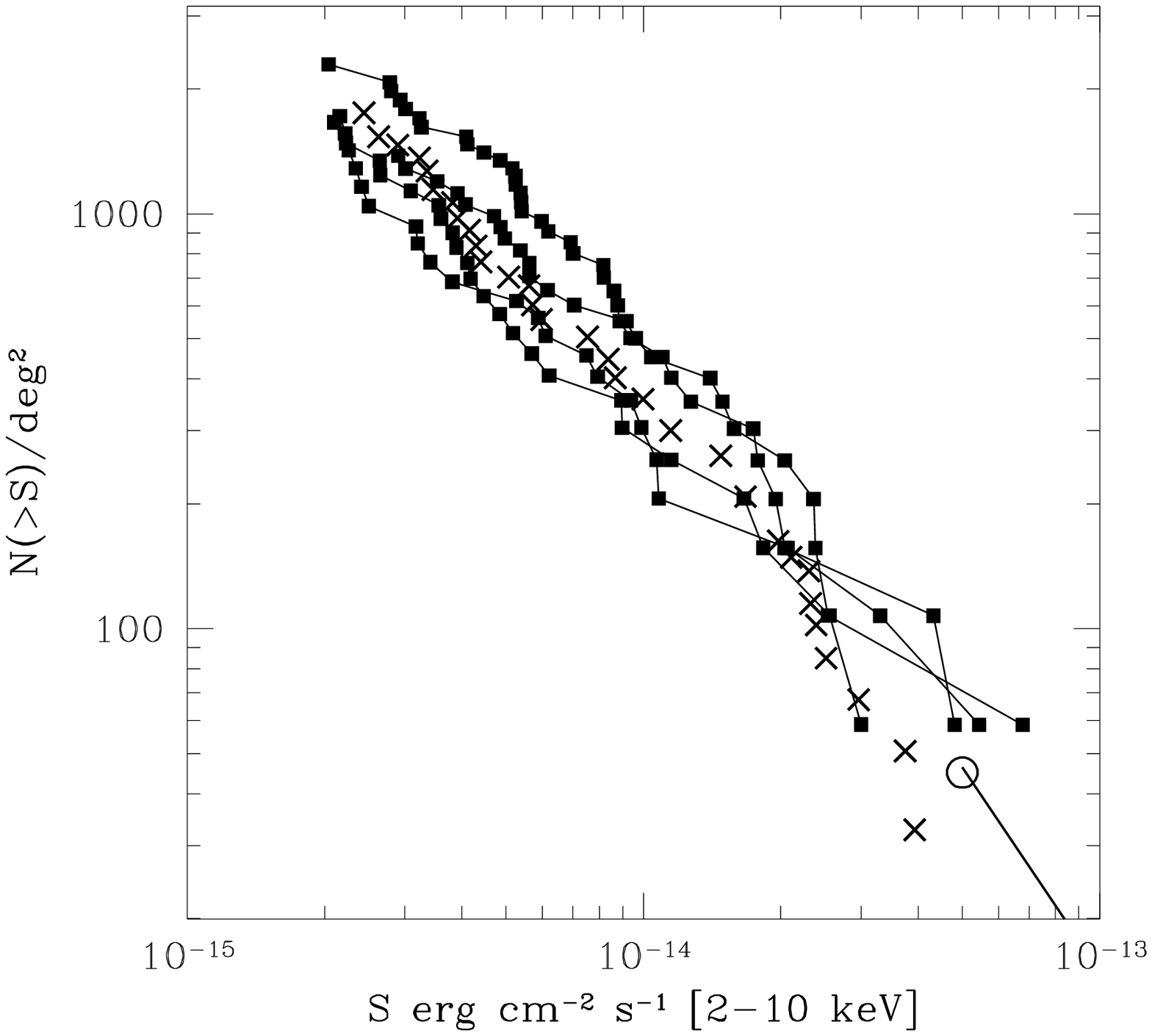}
\caption {Left: Hard counts from the MS--1137 field (filled squares)
compared to the number counts in the Hawaii DF (crosses, Mushotzky et
al. 2000) and the corresponding likelyhood contours in the inset (see
also Figure \ref{fig1}).  Right: hard number counts in the MS--1137
field for each of the 4 ACIS--I chips (filled squares) compared to the
Hawaii DF counts.}
\label{fig3}
\end{figure}

In the CDFS the normalization $K$ (computed at $2\times 10^{-15}$ erg
s$^{-1}$ cm$^{-2}$) for the hard counts is $\sim 40$\% lower than that
in the Hawaii Deep Field (Mushotzky et al. 2000; see the inset of
Figure \ref{fig1}).  The difference is statistically significant
despite the small area of the Hawaii DF (see Paper I and II).
Therefore we investigated if we must revise upwards the fraction of
the resolved XRB in the CDFS.  To understand the nature of this
discrepancy, we analyzed the ACIS--S chip of the Hawaii DF with our
procedure.  We found the same normalization of the number counts,
confirming that the mentioned discrepancy is not due to calibration or
detection technique.

To investigate further this issue and estimate the occurence of this
high normalization, we analyzed two ACIS--I, public Chandra fields,
the Lynx (Stern et al. 2001), and the MS--1137 field.  In the first case
we found perfect agreement with the CDFS results (see Stern et
al. 2001).  In the second case, we found the same ``excess'' of the
0.025 deg$^2$ of the Hawaii DF, as shown in Figure \ref{fig3} (left).
However, the excess is given mainly by two chips where the density of
sources is significantly larger (Figure \ref{fig3}, right).  The
remaining properties of this field are perfectly consistent with the
CDFS, for example the stacked spectrum of the sources is well fitted
by a power law with $\Gamma = 1.39 \pm 0.04$, with the local absorption
fixed to the Galactic value of $1.1\times 10^{20}$ cm$^{-2}$.

Therefore, we find again evidence for positive fluctuations of the
hard sources on $\simeq 10$ arcmin scale.  This positive fluctuations
maybe due to clustering of the hard sources.  It is noteworth to
mention that the same effect is not found in the soft band.  In fact,
we already know that most of the hard sources are in a limited range
of redshifts around 1, while the soft sources span a very large range
(see Paper II), washing out possible effects of clustering.

How much do these positive fluctuations alter our estimate of the
resolved XRB?  To summarize our investigation of Chandra deep fields
to date, we have 0.33 deg$^2$ with a consistent normalization (CDFS,
CDFN, Lynx, Lockman Hole), and 0.089 deg$^2$ with a $\sim 40$\% higher
normalization (Hawaii DF, MS--1137 field).  Considering a 40\% higher
contribution on 20\% of the sky, the resolved contribution in the
investigated flux range ($S<10^{-13}$ er s$^{-12}$ cm$^{-2}$) rises to
($1.42\pm 0.22) \times 10^{-11}$ erg s$^{-1}$ cm$^{-2}$ deg$^{-2}$.
This would bring the total resolved value for $S> 4.5 \times 10^{-16}$
to $(1.67 \pm 0.26) \times 10^{-11}$ erg s$^{-1}$ cm$^{-2}$
deg$^{-2}$.  This value is $\simeq$ 7\% higher than the CDFS value,
but still within the estimated error.

\section{Groups and Clusters}

Another important population of objects in the CDFS, despite its small
solid angle, is constituted by groups and clusters of galaxies.  The
systematic search for extended source found about 20 detections,
including diffuse halos around elliptical galaxies.  Here we focus on
the three brightest extended sources detected at the level of $S\sim
1-3 \times 10^{-15}$ erg s$^{-1}$ cm$^{-2}$ and associated to groups
of galaxies.  We have redshift for the brightest cluster member of one
of them ($z\sim 0.7$), while the optical colors of the others are
consistent with being moderate--redshift poor clusters or groups.  The
stacked spectrum is well fitted by a Raymond--Smith thermal spectrum
with temperature $kT = 1.7^{+0.6}_{-0.4} $ keV and a metallicity of
$Z=0.3 Z_\odot$, after assuming an average redshift of $z\sim 0.5$.
The luminosities are in the range $L\simeq 2-8 \times 10^{42}$ erg
s$^{-1}$ in the soft band.  To this small sample, we can add the three
clusters found in the Lynx field at redshifts 0.57 and 1.26 and 1.27
(Holden et al. 2001, Stanford et al. 2001), and the cluster MS--1137
at $z=0.78$.  If we plot these clusters in the $L$--$T$ plane,
together with other two objects from the literature (see Borgani et
al. 2001), we find that these sources are consistent with the local
$L$--$T$ relation.  This confirms the finding of lack of evolution in
the $L$--$T$ relation (Mushotzky \& Scharf 1997) up to $z\sim 1$.
This behaviour is expected on the basis of pre--heating models for the
ICM (see, e.g., Tozzi \& Norman 2001).

\begin{figure}
\plotone{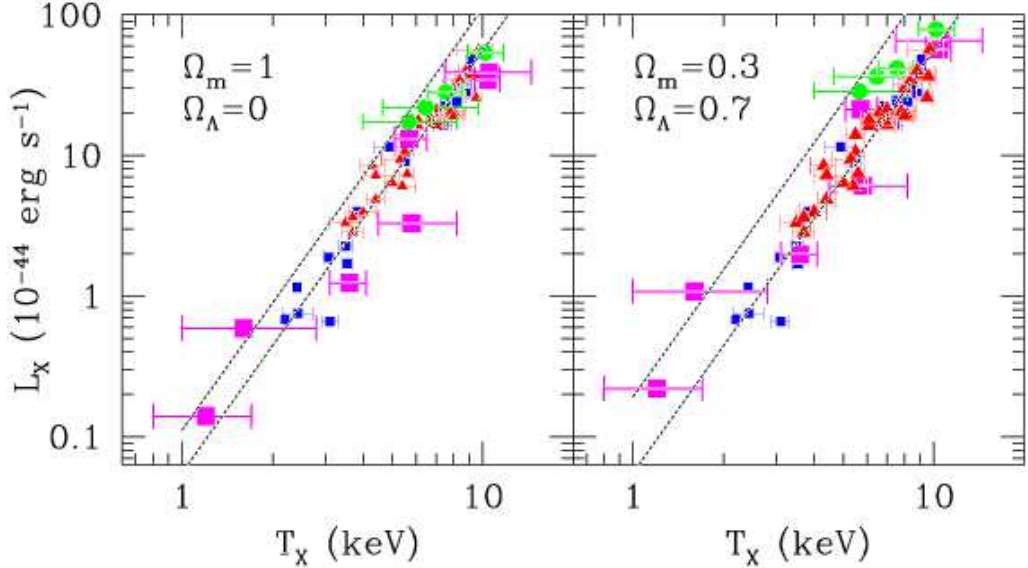}{14cm}
\caption {The Luminosity--Temperature relation for nearby and distant
clusters in two different cosmologies ($h=0.5$).  The nearby clusters
analyzed by Markevitch (1998) and by Arnaud \& Evrard (1999) are
indicated with small triangles and squares respectively.  The large
squares are for the compilation of distant ($0.57\leq z \leq
1.27$) clusters recently observed with Chandra (see Borgani et
al. 2001).  }
\label{fig4}
\end{figure}

\section {Conclusions}

In these Proceedings we reviewed some of the results from the 1Ms
exposure of the CDFS (Rosati et al. 2001).  We reached a flux limit of
$5.5 \times 10^{-17}$ erg s$^{-1}$ cm$^{-2}$ in the 0.5--2 keV soft
band and $4.5 \times 10^{-16}$ erg s$^{-1}$ cm$^{-2}$ in the 2--10 keV
hard band.  For the hard counts, after the inclusion of the ASCA
sources at the bright end, the total contribution to the resolved hard
X--ray background down to our flux limit now amounts to $(1.56 \pm
0.16)\times 10^{-11}$ erg cm$^{-2}$ s$^{-1}$ deg$^{-2}$.  With this
new data we resolved $65-98\%$ of the hard XRB, given the uncertainties
in the unresolved value, which ranges from $1.6$ to $2.4 \times
10^{-11}$ erg s$^{-1}$ cm$^{-2}$ deg$^{-2}$.

The general properties of the X--ray background are now well
established.  We confirm the finding of a progressive hardening of the
sources at lower fluxes, both for the soft and the hard samples.  In
particular, we divided the hard sample into four subsamples and found
average power laws of $\Gamma = 1.68 \pm 0.03$, $\Gamma = 1.42 \pm
0.03$, $\Gamma = 1.10 \pm 0.03$, $\Gamma = 1.05 \pm 0.05$ for the
bright, medium, faint and very faint subsample respectively.  The
progressive hardening at faint fluxes is probably due to a stronger
intrinsic absorption.  The so called ``spectral paradox'' is solved by
a hard, faint population of sources constituted mostly by nearby
($z\leq 1$) absorbed (Type II) AGN with hard luminosities $L\simeq
10^{42}-10^{44}$ erg s$^{-1}$.  Comparison with other deep fields from
Chandra, shows that the AGN found in the CDFS are representative of
the AGN population as a whole.  Some fields have a $\sim 40$\% higher
normalization probably due to clustering of nearby, hard sources.
This excess, averaged over the 6 deep fields mentioned and briefly
reviewed here, can result in an upward revision of the value of the
resolved XRB of about 7\%.

We reviewed also the spectral analysis of the three brightest extended
sources detected in the CDFS.  They are associated with moderate
redshift groups of galaxies ($z\simeq 0.5-0.7$) and their average
spectrum is well fitted by a Raymond--Smith model with a
temperature of $kT = 1.7\pm 0.4$ keV.  This result, added to the
three clusters found in the Lynx field, the cluster MS--1137 and other
two objects from the literature (see Borgani et al. 2001), shows that
there is no hints of evolution of the $L$--$T$ relation up to $z\simeq
1$.  This behaviour is expected in preheating model for the ICM.

The results on AGNs and Clusters reviewed here, give an example of the
high--quality performance of the Chandra satellite.  In particular, the
growing number of deep fields in the Chandra Archive will permit a
detailed exploration of the $z\simeq 1$ Universe in X--rays.


\end{document}